# Unwitting Markowitz' Simplification of Portfolio Random Returns

Victor Olkhov

Independent, Moscow, Russia

victor.olkhov@gmail.com

ORCID: 0000-0003-0944-5113

**Abstract**

In his famous paper, Markowitz (1952) derived the dependence of portfolio random returns on the random returns of its securities. This result allowed Markowitz to obtain his famous expression for portfolio variance. We show that Markowitz's equation for portfolio random returns and the expression for portfolio variance, which results from it, describe a simplified approximation of the real markets when the volumes of all consecutive trades with the securities are assumed to be constant during the averaging interval. To show this, we consider the investor who doesn't trade shares of securities of his portfolio. The investor only observes the trades made in the market with his securities and derives the time series that model the trades with his portfolio as with a single security. These time series describe the portfolio return and variance in exactly the same way as the time series of trades with securities describe their returns and variances. The portfolio time series reveal the dependence of portfolio random returns on the random returns of securities and on the ratio of the random volumes of trades with the securities to the random volumes of trades with the portfolio. If we assume that all volumes of the consecutive trades with securities are constant, obtain Markowitz's equation for the portfolio's random returns. The market-based variance of the portfolio accounts for the effects of random fluctuations of the volumes of the consecutive trades. The use of Markowitz variance may give significantly higher or lower estimates than market-based portfolio variance.

---

This research received no support, specific grant, or financial assistance from funding agencies in the public, commercial, or nonprofit sectors. We welcome offers of substantial support.

# 1. Introduction

In this paper we consider an unintended approximation made by Markowitz (1952) in his classical paper that helped him to derive the expression of the portfolio variance. We don't consider here any portfolio selection problems but study only the derivation of the portfolio variance. Markowitz's expression of the portfolio variance didn't change since 1952, and the only reference needed for our discussion is his famous paper (Markowitz, 1952). We are sure that Markowitz's results are well known and don't need additional explanations. For convenience, we reproduce a few relations from his paper. Markowitz considered the portfolio composed at time $t_0$ of $j=1,...J$ securities. At current time $t$ the mean return $R(t,t_0)$ (1.1) of the unchanged portfolio takes the form (Markowitz, 1952, p. 78):

$$R(t, t_0) = \sum_{j=1}^{J} R_j(t, t_0) X_j(t_0) \qquad (1.1)$$

The functions $R_j(t,t_0)$ denote the average returns of the securities $j$ at current time $t$ with respect to time $t_0$ in the past. The variables $X_j(t_0)$ in (1.1) denote the relative amounts invested into securities $j$ at time $t_0$. All prices are adjusted to the current time $t$. Markowitz concluded that the random return $R$ (1.2) of the portfolio has a linear form similar to (1.1):

$$R = \sum_{j=1}^{J} R_j X_j(t_0) \qquad (1.2)$$

Markowitz highlighted (Markowitz, 1952, p. 81) that "*the $R_j$ (and consequently R) are considered to be random variables.*" The relation (1.2) is a key statement made by Markowitz. On the one hand, (1.2) justifies the expression of the mean return $R(t,t_0)$ (1.1), and on the other hand, (Markowitz, 1952, p. 81) directly leads to the expression the portfolio variance $\Theta_M(t,t_0)$ (1.3) as a quadratic form in the variables $X_j(t_0)$:

$$\Theta_M(t, t_0) = \sum_{j,k=1}^{J} \theta_{jk}(t, t_0) X_j(t_0) X_k(t_0) \qquad (1.3)$$

The coefficients $\theta_{jk}(t,t_0)$ (1.4) are equal to the covariance of securities $j$ and $k$ of the portfolio:

$$\theta_{jk}(t, t_0) = E\big[\big(R_j(t_i, t_0) - E[R_j(t_i, t_0)]\big)\big(R_k(t_i, t_0) - E[R_k(t_i, t_0)]\big)\big] \qquad (1.4)$$

The "obvious" dependence (1.2) of a random return $R$ of the unchanged portfolio on the random returns $R_j$ of its securities served as a reliable tool for the justification of the portfolio variance $\Theta_M(t,t_0)$ (1.3) since 1952.

However, the devil is in the details. We show that while making a short transition from the description of the mean return $R(t,t_0)$ (1.1) to the description of random returns (1.2), Markowitz unintentionally made an important and non-trivial assumption. However, his hidden and unwilled assumption describes a rather limited model of real trades made in the markets with the securities. Left unnoticed since 1952, it probably limited the description of real markets that reveal highly irregular or random fluctuations of volumes of trades.

## 2. Unintentional Markowitz's approximation

We follow Markowitz and consider the investor who collected his portfolio with shares of $j=1,...J$ securities in the past at time $t_0$ and since then doesn't trade shares of his portfolio. Now, let us assume that the investor wants to estimate the return and variance of the portfolio at the current time $t$. To do that, the investor needs not to trade shares of his portfolio. The investor may observe the time series of the current trades made in the market with the shares of his securities during some averaging interval. We show how that allows the investor to derive the time series that model the trades with his portfolio as with a single security and to estimate the current return and variance of his unchanged portfolio.

The time series of trades with the securities $j=1,...J$ of the portfolio give the evident justification of the mean return $R(t,t_0)$ (1.1). The mean contribution during the averaging interval of the security $j$ to the portfolio value at time $t$ equals the average return $R_j(t,t_0)$ of security $j$ multiplied by the initial value $C_j(t_0)$ of the shares of security $j$ in the portfolio at time $t_0$ in the past. The current value $Q_\Sigma(t)$ (2.1) of the portfolio equals the sum of the contributions of all securities:

$$Q_\Sigma(t) = \sum_{j=1}^{J} R_j(t,t_0) C_j(t_0) \quad ; \quad Q_\Sigma(t_0) = \sum_{j=1}^{J} C_j(t_0) \quad ; \quad X_j(t_0) = \frac{C_j(t_0)}{Q_\Sigma(t_0)} \quad (2.1)$$

The initial value $Q_\Sigma(t_0)$ (2.1) of the portfolio at time $t_0$ is a sum of the initial values or amounts $C_j(t_0)$ invested into securities $j$. The ratio of the current value $Q_\Sigma(t)$ to the initial value $Q_\Sigma(t_0)$ (2.1) of the portfolio gives the mean portfolio return $R(t,t_0)$ (1.1). The relative amounts $X_j(t_0)$ invested into security $j$ at time $t_0$ are determined in (2.1).

However, the justification of the equation (1.2) that describes a random return $R$ of the portfolio as a linear form (1.2) of random returns $R_j$ of securities $j=1,...J$ similar to a linear dependence of the mean return $R(t,t_0)$ (1.1) is not evident, and moreover, is valid only for a rather limited approximation of the change of the volumes of the consecutive market trades. As we show below, the equation (1.2) derived by Markowitz, is correct only if all volumes of the consecutive trades with all securities $j=1,...J$ of the portfolio are assumed constant during the averaging interval.

To explain the economic meaning of Markowitz's involuntary approximation, let us consider the time series of the values $C_j(t_i)$, volumes $U_j(t_i)$, and prices $p_j(t_i)$ of the consecutive trades made in the market with securities $j=1,...J$ of the portfolio at time $t_i$ during the averaging interval. We assume that the consecutive trades with securities are made at time $t_i$ with a small constant span $\varepsilon$ between the trades, so $t_{i+1}=t_i+\varepsilon$, and hence, the averaging interval $\Delta$ (2.2) contains only a finite number $N$ of trades with each security $j$:

$$\Delta = \left[t - \frac{\Delta}{2}; t + \frac{\Delta}{2}\right] \quad ; \quad t_i \in \Delta \quad ; \quad i = 1, \ldots N \quad ; \quad N \cdot \varepsilon = \Delta \qquad (2.2)$$

A finite number $N$ of terms of time series define the approximation of the mean value $C(t)$ and volume $U(t)$ of trades at time $t$:

$$C_j(t) = \frac{1}{N}\sum_{i=1}^{N} C_j(t_i) \quad ; \quad C_{\Sigma j}(t) = N \cdot C_j(t) \quad ; \quad U_j(t) = \frac{1}{N}\sum_{i=1}^{N} U_j(t_i) \quad ; \quad U_{\Sigma j}(t) = N \cdot U_j(t) \qquad (2.3)$$

The functions $C_{\Sigma j}(t)$ and $U_{\Sigma j}(t)$ in (2.3) denote the total value and volume of trades with security $j$ during $\Delta$ (2.2). The values $C_j(t_i)$, volumes $U_j(t_i)$, and prices $p_j(t_i)$ follow the equation:

$$C_j(t_i) = p_j(t_i) \cdot U_j(t_i) \qquad (2.4)$$

We recall that the investor doesn't trade his shares, and his portfolio remains unchanged. The investor observes the time series of the values $C_j(t_i)$, volumes $U_j(t_i)$, and prices $p_j(t_i)$ of the trades with securities $j=1,\ldots J$ of his portfolio made in the market during $\Delta$ (2.2). These trades don't change the numbers of shares of the portfolio. Relations (2.3; 2.4) determine the mean price $p_j(t)$ (2.5) of security $j$ during $\Delta$ (2.2):

$$p_j(t) = \frac{C_{\Sigma j}(t)}{U_{\Sigma j}(t)} = \frac{C_j(t)}{U_j(t)} = \frac{1}{U_{\Sigma j}(t)} \sum_{i=1}^{N} p_j(t_i) U_j(t_i) \qquad (2.5)$$

The mean price $p_j(t)$ (2.5) has the form of volume weighted average price (VWAP) (Berkowitz et al., 1988; Duffie and Dworczak, 2021). We give these simple definitions to determine the mean $R_j(t,t_0)$ and the random $R_j(t_i,t_0)$ returns of security $j$:

$$R_j(t_i, t_0) = \frac{p_j(t_i)}{p_j(t_0)} \quad ; \quad R_j(t, t_0) = \frac{p_j(t)}{p_j(t_0)} = \frac{1}{U_{\Sigma j}(t)} \sum_{i=1}^{N} R_j(t_i, t_0) U_j(t_i) \qquad (2.6)$$

In (2.6), functions $p_j(t_0)$ determine the price of shares of security $j$ of the portfolio at time $t_0$ in the past. For convenience, we consider the "gross" returns (2.6) instead of the "usual" definition of returns $r_j(t,t_0) = R_j(t,t_0)-1$. Both definitions have the same variances.

The random return $R_j(t_i,t_0)$ (2.6) of security $j$ is determined by the random price $p_j(t_i)$ of trade with security $j$ made at time $t_i$ during $\Delta$ (2.2). The mean return $R_j(t,t_0)$ (2.6) of security $j$ is determined by the VWAP $p_j(t)$ (2.5). One may call the definition (2.6) as volume weighted average returns $R_j(t,t_0)$. The returns of the security $j$ are determined by the time series (2.3-2.6) of trades with security $j$ during $\Delta$ (2.2). To define the random returns $R(t_i,t_0)$ of the portfolio in the form that is similar to (2.6), we introduce the time series of trades with the portfolio.

## 3. Time series of trades with the portfolio

At first, let us specify the portfolio that was composed by the investor at time $t_0$ in the past and since then the numbers $U_j(t_0)$ of shares of its securities $j=1,\ldots J$ remain unchanged. The values

$C_j(t_0)$ and prices $p_j(t_0)$ of the shares of security $j$ at time $t_0$ follow an equation (3.1):

$$C_j(t_0) = p_j(t_0) \cdot U_j(t_0) \tag{3.1}$$

We introduce the total number of shares $W_\Sigma(t_0)$ of all securities of the portfolio and its total value $Q_\Sigma(t_0)$ as the sums of the numbers $U_j(t_0)$ and values $C_j(t_0)$ of securities $j=1,...J$:

$$Q_\Sigma(t_0) = \sum_{j=1}^{J} C_j(t_0) \quad ; \quad W_\Sigma(t_0) = \sum_{j=1}^{J} U_j(t_0) \tag{3.2}$$

The total value $Q_\Sigma(t_0)$ and the total number of shares $W_\Sigma(t_0)$ of the portfolio determine the price $s(t_0)$ (3.3) per share of the portfolio at time $t_0$:

$$Q_\Sigma(t_0) = s(t_0) W_\Sigma(t_0) \quad ; \quad s(t_0) = \sum_{j=1}^{J} p_j(t_0) x_j(t_0) \quad ; \quad x_j(t_0) = \frac{U_j(t_0)}{W_\Sigma(t_0)} \tag{3.3}$$

The functions $x_j(t_0)$ have the meaning of relative numbers of the shares of security $j$ in the portfolio. The number of shares $W_\Sigma(t_0)$ of the portfolio remains constant, but the value $Q_\Sigma(t)$ of the portfolio at the current time $t$ during $\Delta$ (2.2) depends on the current mean prices $p_j(t)$ (2.5) of trades with the securities during $\Delta$ (2.2). To estimate the current value $Q_\Sigma(t)$ of the portfolio during $\Delta$ (2.2) using the time series of the values $C_j(t_i)$, volumes $U_j(t_i)$, and prices $p_j(t_i)$ of the trades with securities $j=1,...J$, let us for each security $j$ define the factor $\lambda_j$ (3.4) that is equal to the ratio of the number of shares $U_j(t_0)$ (3.1) of security $j$ in the portfolio at time $t_0$ to the current total volume $U_{\Sigma j}(t)$ (2.3) of trades with security $j$ made in the market during $\Delta$ (2.2):

$$\lambda_j = \frac{U_j(t_0)}{U_{\Sigma j}(t)} \tag{3.4}$$

We highlight that the equation (2.4) has the important property that the multiplication of the values $C_j(t_i)$ and volumes $U_j(t_i)$ by a constant factor $\lambda_j$ doesn't change the price $p_j(t_i)$. Let us use this and define normalized values $c_j(t_i)$ and volumes $u_j(t_i)$ (3.5) of trades with security $j$:

$$c_j(t_i) = \lambda_j \cdot C_j(t_i) \quad ; \quad u_j(t_i) = \lambda_j \cdot U_j(t_i) \tag{3.5}$$

The equations (3.4; 3.5) transform (2.4) into (3.6):

$$c_j(t_i) = p_j(t_i) u_j(t_i) \quad or \quad \lambda_j \cdot C_j(t_i) = p_j(t_i) \cdot \lambda_j \cdot U_j(t_i) \tag{3.6}$$

From (3.4; 3.5), obtain that the total normalized volume $u_{\Sigma j}(t)$ (2.3; 3.7) of trades during $\Delta$ (2.2) exactly equals the number of shares $U_j(t_0)$ (3.1) of security $j$ in the portfolio:

$$u_{\Sigma j}(t) = \sum_{i=1}^{N} u_j(t_i) = \frac{U_j(t_0)}{U_{\Sigma j}(t)} \sum_{i=1}^{N} U_j(t_i) = U_j(t_0) \quad ; \quad u_j(t) = \frac{u_{\Sigma j}(t)}{N} \tag{3.7}$$

The time series of the normalized values $c_j(t_i)$ and volumes $u_j(t_i)$ (3.5) of trades with securities $j=1,...J$ during $\Delta$ (2.2) model the trades with securities of the portfolio and their total normalized volumes $u_{\Sigma j}(t)=U_j(t_0)$ (3.7) exactly equal to the number of shares $U_j(t_0)$ of each security. That determines the time series of the values $Q(t_i)$ and volumes $W(t_i)$ of trades with the portfolio:

$$Q(t_i) = \sum_{j=1}^{J} c_j(t_i) \quad ; \quad W(t_i) = \sum_{j=1}^{J} u_j(t_i) \tag{3.8}$$

The sum of normalized values $c_j(t_i)$ of trades with all securities $j=1,...J$ determines the value $Q(t_i)$ (3.8) of the trade with the portfolio at time $t_i$. The sum of normalized volumes $u_j(t_i)$ of trades with all securities $j=1,...J$ determines the volume $W(t_i)$ (3.8) of the trade with the portfolio at time $t_i$. The total sum $W_\Sigma(t)$ (3.9) of the volumes of trades with the portfolio during $\Delta$ (2.2) equals the total number of shares of the portfolio $W_\Sigma(t_0)$ (3.2):

$$W_\Sigma(t) = \sum_{i=1}^{N} W(t_i) = \sum_{j=1}^{J} \sum_{i=1}^{N} u_j(t_i) = \sum_{j=1}^{J} U_j(t_0) = W_\Sigma(t_0) \qquad (3.9)$$

Thus, the time series of the values $Q(t_i)$ and volumes $W(t_i)$ describe the trades with the portfolio as with a single security in the same way as the time series of the values $C_j(t_i)$ and volumes $U_j(t_i)$ describe the trades with security $j$. Similar to equation (2.4), we introduce the price $s(t_i)$ (3.10) of the trade with the portfolio as with a single security at time $t_i$ during $\Delta$ (2.2):

$$Q(t_i) = s(t_i)\, W(t_i) \quad ; \quad t_i \in \Delta \quad ; \quad i = 1, \dots N \qquad (3.10)$$

The total current value $Q_\Sigma(t)$ (3.11) of trades with all $W_\Sigma(t_0)$ (3.2) shares of the portfolio or the current value $Q_\Sigma(t)$ of the portfolio during $\Delta$ (2.2) equals:

$$Q_\Sigma(t) = s(t) W_\Sigma(t) = \sum_{i=1}^{N} Q(t_i) = \sum_{j=1}^{J} \sum_{i=1}^{N} c_j(t_i) = \sum_{j=1}^{J} \sum_{i=1}^{N} p_j(t_i) u_j(t_i) \qquad (3.11)$$

The equations (3.11) at time $t$ determine the mean price $s(t)$ (3.12) per share of the portfolio during $\Delta$ (2.2) in the form that is similar to VWAP $p_j(t)$ (2.25):

$$s(t) = \frac{Q_\Sigma(t)}{W_\Sigma(t)} = \frac{1}{W_\Sigma(t_0)} \sum_{i=1}^{N} s(t_i)\, W(t_i) \qquad (3.12)$$

## 4. Mean and random portfolio returns

The equations (3.3; 3.10; 3.12) determine the mean $R(t,t_0)$ and the random $R(t_i,t_0)$ returns of the portfolio during $\Delta$ (2.2) in a form that is alike to the definitions (2.6) of the mean and the random returns of security $j$:

$$R(t_i, t_0) = \frac{s(t_i)}{s(t_0)} \quad ; \quad R(t, t_0) = \frac{s(t)}{s(t_0)} = \frac{1}{W_\Sigma(t)} \sum_{i=1}^{N} R(t_i, t_0)\, W(t_i) \qquad (4.1)$$

From (2.5; 3.4; 3.6; 3.8; 3.9; 3.11; 3.12), obtain the decomposition of the mean price $s(t)$ (4.2) of the portfolio:

$$s(t) = \frac{1}{W_\Sigma(t_0)} \sum_{j=1}^{J} \sum_{i=1}^{N} p_j(t_i) u_j(t_i) = \frac{1}{W_\Sigma(t_0)} \sum_{j=1}^{J} \frac{U_j(t_0)}{U_{\Sigma j}(t)} \sum_{i=1}^{N} p_j(t_i) U_j(t_i)$$

$$s(t) = \sum_{j=1}^{J} p_j(t) \frac{U_j(t_0)}{W_\Sigma(t_0)} = \sum_{j=1}^{J} p_j(t)\, x_j(t_0) \quad ; \quad x_j(t_0) = \frac{U_j(t_0)}{W_\Sigma(t_0)} \qquad (4.2)$$

We recall that $x_j(t_0)$ (3.3; 4.2) denotes the relative numbers of shares of securities $j$ in the portfolio. The decomposition of the mean price $s(t)$ (3.12; 4.2) of the portfolio by the mean prices $p_j(t)$ (2.5) of its securities determines the decomposition of the mean return $R(t,t_0)$ (4.1) by the mean returns $R_j(t,t_0)$ (2.6):

$$R(t,t_0) = \frac{s(t)}{s(t_0)} = \frac{1}{s(t_0)}\sum_{j=1}^{J} p_j(t)\, x_j(t_0) = \sum_{j=1}^{J} \frac{p_j(t)}{p_j(t_0)} \frac{p_j(t_0)}{s(t_0)} \frac{U_j(t_0)}{W_\Sigma(t_0)}$$

From (2.1; 3.1; 3.3), obtain the mean return $R(t,t_0)$ (4.3) that coincides with (1.1):

$$R(t,t_0) = \sum_{j=1}^{J} R_j(t,t_0)\, X_j(t_0) \quad ; \quad X_j(t_0) = \frac{C_j(t_0)}{Q_\Sigma(t_0)} = \frac{p_j(t_0)\cdot U_j(t_0)}{s(t_0)\cdot W_\Sigma(t_0)} \tag{4.3}$$

The time series of the values $Q(t_i)$ and volumes $W(t_i)$ (3.8-3.12) that describe the trades with the portfolio as with a single security give the additional proof of the decomposition (1.1; 4.3) of mean return $R(t,t_0)$ of the portfolio by the mean returns $R_j(t,t_0)$ of its securities.

However, the time series of the values $Q(t_i)$ and volumes $W(t_i)$ (3.8-3.12) reveal that the decomposition of the random return $R(t_i,t_0)$ (4.1) of the portfolio differs from Markowitz's equation (1.2). From (4.1; 3.10), obtain:

$$R(t_i,t_0) = \frac{s(t_i)}{s(t_0)} = \frac{Q(t_i)}{s(t_0)W(t_i)} = \frac{1}{s(t_0)W(t_i)}\sum_{j=1}^{J} C_j(t_i) = \frac{1}{s(t_0)W(t_i)}\sum_{j=1}^{J} p_j(t_i)\, u_j(t_i)$$

Simple transformations give:

$$R(t_i,t_0) = \sum_{j=1}^{J} \frac{p_j(t_i)}{p_j(t_0)} \frac{p_j(t_0)U_j(t_0)}{s(t_0)W_\Sigma(t_0)} \frac{u_j(t_i)}{W(t_i)} \frac{W_\Sigma(t_0)}{U_j(t_0)} = \sum_{j=1}^{J} R_j(t_i,t_0)\, X_j(t_0) \frac{u_j(t_i)}{W(t_i)} \frac{W_\Sigma(t_0)}{U_j(t_0)}$$

Finally, obtain the equation of the portfolio random return $R(t_i,t_0)$ (4.4) that accounts for the impact of fluctuations of the volumes of the consecutive trades, which was missed in (1.2):

$$R(t_i,t_0) = \sum_{j=1}^{J} R_j(t_i,t_0) \cdot \frac{x_j(t_i)}{x_j(t_0)} \cdot X_j(t_0) \quad ; \quad x_j(t_i) = \frac{u_j(t_i)}{W(t_i)} = \frac{U_j(t_0)}{U_{\Sigma j}(t)} \cdot \frac{U_j(t_i)}{W(t_i)} \tag{4.4}$$

The random relative volumes $x_j(t_i)$ (4.4) at time $t_i$ equal to the ratio of the random volumes $u_j(t_i)$ (3.5) of the normalized trades with securities $j$ to the random volumes $W(t_i)$ (3.8) of trades with the portfolio. The decomposition of the random return $R(t_i,t_0)$ (4.1; 4.4) of the portfolio by the random returns $R_j(t_i,t_0)$ (2.6) of its securities depends on the random relative volumes $x_j(t_i)$ (4.4) that are additional to Markowitz's expression of random returns (1.2). The differences between the market-based expression of the random return $R(t_i,t_0)$ (4.4) and Markowitz's expression (4.3) reflect the different approximations of the values of the consecutive trades. These distinctions vanish and make no impact while one evaluates the mean return $R(t,t_0)$ (4.3) of the portfolio. However, the assessment of the portfolio variance via (4.4) results in significant distinctions from Markowitz variance (1.3). The economic nature of the distinctions between Markowitz's assessment of the random return of the portfolio (1.2) and (4.4) is rather simple. One can easily show that if the volumes $U_j(t_i)$ of all $N$ consecutive trades with all securities $j=1,...J$ of the portfolio are assumed constant during the averaging interval $\Delta$ (2.2), then:

$$U_j(t_i) = U_j = \frac{U_{\Sigma j}(t)}{N} \; ; \; u_j(t_i) = u_j = \frac{U_j(t_0)}{U_{\Sigma j}(t)} \frac{U_{\Sigma j}(t)}{N} = \frac{U_j(t_0)}{N} \; ; \; W(t_i) = W = \frac{W_\Sigma(t_0)}{N}$$

$$x_j(t_i) = \frac{u_j}{W} = \frac{U_j(t_0)}{W_\Sigma(t_0)} = x_j(t_0) \tag{4.5}$$

Thus, if one assumes that all volumes $U_j(t_i)$ of the consecutive trades are constant, the equation (4.4) on the random probability $R(t_i,t_0)$ of the portfolio takes the form (1.2) that was proposed by Markowitz (1952). That highlights the origin of Markowitz variance $\Theta_M(t,t_0)$ (1.3) that describes the approximation when all volumes $U_j(t_i)$ of the consecutive trades with securities of the portfolio are assumed constant. This approximation greatly simplifies the assessment of the portfolio variance. Since 1952 it has given a strong impetus for the successful development of the optimal portfolio selection models and portfolio theory as well. However, one should keep in mind the existing limitations of Markowitz's results that completely ignore the random fluctuations of the trade volumes of real markets.

## 5. Market-based portfolio variance

The time series of the values $Q(t_i)$, volumes $W(t_i)$ (3.8), and prices $s(t_i)$ (3.10) of trades with the portfolio determine the random return $R(t_i,t_0)$ (4.4) and the market-based portfolio variance that accounts for the impact of fluctuations of the volumes of the consecutive trades. For the readers' convenience, we present the expressions of market-based portfolio variance. One can find the derivation and clarification of these results in Olkhov (2025a; 2025b).

At current time $t$, market-based portfolio variance $\Theta(t,t_0)$ (5.1) with respect to time $t_0$ in the past takes the form:

$$\Theta(t,t_0) = \frac{\psi^2(t) - 2\varphi(t) + \chi^2(t)}{1 + \chi^2(t)} R^2(t,t_0) \tag{5.1}$$

The variance $\Theta(t,t_0)$ (5.1) depends on the mean return $R(t,t_0)$ (4.3) and on the coefficients of variation of the trade values $\psi(t)$ (5.2) and of trade volumes $\chi(t)$ (5.2) averaged during $\Delta$ (2.2):

$$\psi^2(t) = \frac{cov\{Q(t),Q(t)\}}{Q^2(t;1)} = \frac{\Psi_Q(t)}{Q^2(t;1)} \quad ; \quad \chi^2(t) = \frac{cov\{W(t),W(t)\}}{W^2(t;1)} = \frac{\Psi_W(t)}{W^2(t;1)} \tag{5.2}$$

The function $\varphi(t)$ in (5.1) denotes the ratio (5.3) of the covariance of values $Q(t_i)$ and volumes $W(t_i)$ of the portfolio trades to their mean values $Q(t;1)$ (5.5) and $W(t;1)$ (5.7).

$$\varphi(t) = \frac{cov\{Q(t),W(t)\}}{Q(t;1)W(t;1)} \quad ; \quad cov\{Q(t),W(t)\} = \frac{1}{N}\sum_{i=1}^{N}(Q(t_i) - Q(t;1))(W(t_i) - W(t;1)) \tag{5.3}$$

The functions $\Psi_Q(t)$ (5.4) and $\Psi_W(t)$ (5.6) denote the square of standard deviations of the values $Q(t_i)$ and volumes $W(t_i)$ of trades:

$$\Psi_Q(t) = cov\{Q(t),Q(t)\} = \frac{1}{N}\sum_{i=1}^{N}(Q(t_i) - Q(t;1))^2 = Q(t;2) - Q^2(t;1) \tag{5.4}$$

The functions $Q(t;1)$ and $Q(t;2)$ (5.5) denote the mean values and the mean squares of values:

$$Q(t;1) = \frac{1}{N}\sum_{i=1}^{N}Q(t_i) \quad ; \quad Q(t;2) = \frac{1}{N}\sum_{i=1}^{N}Q^2(t_i) \tag{5.5}$$

$$\Psi_W(t) = cov\{W(t),W(t)\} = \frac{1}{N}\sum_{i=1}^{N}(W(t_i) - W(t;1))^2 = W(t;2) - W^2(t;1) \tag{5.6}$$

The functions $W(t;1)$, $W(t;2)$ (5.7) denote the mean volumes and the mean squares of volumes.

$$W(t;1) = \frac{1}{N}\sum_{i=1}^{N} W(t_i) \quad ; \quad W(t;2) = \frac{1}{N}\sum_{i=1}^{N} W^2(t_i) = W^2(t;1)[1 + \chi^2(t)] \qquad (5.7)$$

To simplify the assessments of market-based variance $\Theta(t,t_0)$, (5.1) Olkhov (2025b) derived the Taylor expansion (5.8) of the variance $\Theta(t,t_0)$ (5.1) up to the 2$^{nd}$ degree of the coefficient of variation $\chi(t)$ (5.2) of the trade volumes, taking Markowitz variance $\Theta_M(t,t_0)$ (1.3) as a zero approximation for $\chi(t)=0$.

$$\Theta(t,t_0) = \Theta_M(t,t_0) - 2a\, \Theta_M^{\frac{1}{2}}(t,t_0)\, R(t,t_0)\, \chi(t) + [R^2(t,t_0) - \Theta_M(t,t_0)]\, \chi^2(t) \qquad (5.8)$$

For three limiting cases with high and low fluctuations of returns of the portfolio and with zero covariance of trade values and volumes, Olkhov (2025b) revealed that the impact of fluctuations of trade volumes causes that Markowitz variance $\Theta_M(t,t_0)$ (1.3) may significantly overestimate or greatly underestimate the market-based variance of the portfolio that accounts for the impact of fluctuations of the volumes of consecutive trades.

The decomposition of the portfolio variance $\Theta(t,t_0)$ (5.1) by the variances of its securities and the Taylor series of the decomposition by the coefficients of variations $\chi_j(t)$ of trade volumes of securities $j=1,...J$ are given in Olkhov (2025a; 2025b).

## 6. Conclusion

The portfolio selection and the portfolio theory as a whole relies heavily on the dependence of portfolio random returns on the random returns of its securities. To a great extent this dependence serves as a basis for the assessments of the impact of the risks of securities on portfolio risks. However, any "intuition" or implicit simplification, like one made by Markowitz to describe a linear dependence (1.2) of the portfolio risks on risks of its securities, has a particular economic ground and limitations. The reliable forecasts of modern financial markets and the secure projections of the mean and variance of returns essentially depend on the economic approximations made to model the market trades. Markowitz's unintended approximation of the volumes of the consecutive trades with securities as being constant gave a simple and useful expression of the variance that boosted the development of the optimal portfolio selection.

Left unnoticed for almost 75 years, this simplified approximation probably might have limited in some sense the description of the random markets. One should care that the use of Markowitz variance (1.3) may give much lower or greatly higher estimates than the market-based portfolio variance, which accounts for the impact of fluctuations of the volumes of consecutive trades (Olkhov, 2025a; 2025b). The investors and portfolio managers should keep that in mind.

# References

Berkowitz, S., Logue, D. and E. Noser, Jr., (1988), The Total Cost of Transactions on the NYSE, *The Journal of Finance*, 43, (1), 97-112

Duffie, D. and P. Dworczak, (2021), Robust Benchmark Design, *Journal of Financial Economics*, 142(2), 775–802

Markowitz, H., (1952), Portfolio Selection, *Journal of Finance*, 7(1), 77-91

Olkhov, V., (2025a), Market-Based Portfolio Variance, *SSRN WPS* 5212636, 1-17

Olkhov, V., (2025b), Markowitz Variance Can Vastly Undervalue or Overestimate Portfolio Variance and Risks, *SSRN WPS* 5370800, 1-20